\documentclass[journal]{IEEEtran}
\ifCLASSINFOpdf
  \usepackage[pdftex]{graphicx}
  \DeclareGraphicsExtensions{.pdf,.jpeg,.png}
\else
\fi
\usepackage[cmex10]{amsmath}
\usepackage{array}

\usepackage{stfloats}

\newcounter{MYtempeqncnt}
\begin{document}
%
\title{Performance Indicator for MIMO MMSE Receivers in the Presence of Channel Estimation Error}
%
%
%
\author{Eren Eraslan, \IEEEmembership{Student Member, IEEE}, Babak Daneshrad, \IEEEmembership{Member, IEEE}, and Chung-Yu Lou
\thanks{The authors are with the Department of Electrical Engineering, University
of California, Los Angeles, 56-125B Engineering IV Building, Los
Angeles, CA 90095-1594, USA (emails: \{eren, babak, cylou\}@ee.ucla.edu).}}

\newcommand{\tr}{\mbox{tr}}

\maketitle

\begin{abstract}
We present the derivation of post-processing SNR for Minimum-Mean-Squared-Error (MMSE) receivers with imperfect channel estimates, and show that it is an accurate indicator of the error rate performance of MIMO systems in the presence of channel estimation error. Simulation results show the tightness of the analysis.
\end{abstract}


%
\IEEEpeerreviewmaketitle

\section{Introduction}
%
%
%
%
\IEEEPARstart{T}{he} key component of a Multiple-Input-Multiple-Output (MIMO) communication system in terms of performance and complexity is the MIMO detector, which is used for separating independent data streams at the receiver. The Maximum Likelihood (ML) detectors achieve the optimal error rate performance. However, these types of detectors, including the near-optimal sphere decoder and its variants, are usually not suitable for practical systems due to their high complexity. Linear detectors, such as Zero-Forcing (ZF) and MMSE, achieve suboptimal performance, however, they are widely used in practical systems due to their low complexity implementations. Among linear receivers, MMSE is the optimal solution and seems to be the mainstream implementation choice due to its superior performance over ZF detectors.

Perfect channel state information (CSI) is usually assumed in the literature when simulating or analyzing the performance of linear detectors \cite{pingli,kim}. However, in practice the channel estimates are inherently noisy. Important work \cite{wang}, \cite{iranli} has characterized the error rate performance of ZF receivers in the presence of channel estimation error. Nevertheless, less is known for the case of MMSE detectors in practical scenarios. For ZF and MMSE receivers, the joint effect of phase noise and channel estimation error is considered in \cite{garcia} and the performance is analyzed in terms of the degradation in signal-to-noise-plus-interference-ratio (SINR) without expressing the closed form performance indicators or error rate analysis. The SINR derivations for the MMSE case in \cite{garcia} are done only for low SNR region. In both \cite{wang} and \cite{garcia}, channel estimation error variance is assumed to be constant for all SNRs. This is not realistic approach for packet based or bursty communication systems as the channel estimation error is in fact a function of the SNR. In this letter, we analyze the MMSE receivers in the presence of channel estimation error, and derive a closed form post-processing SNR expression, which provides an accurate estimate of the error rate performance. The error rate performance is investigated for both the constant channel estimation error variance case and the case with a realistic channel estimation algorithm where the estimation error variance is clearly dependent on the channel SNR. We believe that it is a very useful tool for throughput prediction in link adaptation protocols and for error rate analysis in general. Accuracy of the analytical results is verified through simulations.

\section{System Model and Derivations}
We consider a MIMO system where the transmitter is equipped with $N_t$ antennas, and the receiver uses $N_r$ antennas. The $N_r \times 1$ received signal vector $\mathbf{y}$ can be expressed as
\begin{equation}
\label{hxn}
	\mathbf{y}=\mathbf{H}\mathbf{x}+\mathbf{n}
\end{equation}
where $\mathbf{x}$ is the transmitted signal vector, $\mathbf{H}$ is the $N_r \times N_t$ channel matrix, and $\mathbf{n}$ is the $N_r \times 1$ additive Gaussian noise vector with zero mean and covariance matrix $E\left[\mathbf{n}\mathbf{n}^H\right]=N_0\mathbf{I}$. We assume an uncorrelated Rayleigh flat channel, i.e. entries of $\mathbf{H}$ are i.i.d. zero mean circularly symmetric complex Gaussians (ZMCSCG) with unit variance, and the signal energy at each transmit antenna is assumed to be equal to $E_s$.

The receiver can estimate the transmitted signal vector by applying the MMSE detector to the received signal, $\hat{\mathbf{x}}=\mathbf{W}\mathbf{y}=\mathbf{W}\mathbf{H}\mathbf{x}+\mathbf{W}\mathbf{n}$. Using the orthogonality principle \cite{sayed}, the MMSE detector $\mathbf{W}$ is derived as 
\begin{equation}
\label{mmse}
	\mathbf{W}=\left[\mathbf{H}^{H}\mathbf{H}+\frac{N_0}{E_s}\mathbf{I}\right]^{-1}\mathbf{H}^{H}
\end{equation}

At the output of the MMSE detector, the residual signal plus interference from other spatial streams is well approximated as Gaussian \cite{peng} and the post-processing SNR (PPSNR) of $k^{th}$ spatial stream is calculated as\footnote{$\left(\ldots\right)_{k,l}$ denotes the $\left(k,l\right)^{th}$ entry of the matrix.}
\begin{equation}
\label{ppsnr}
	\gamma_k=\frac{E_s\left|\left(\mathbf{WH}\right)_{k,k}\right|^2}{E_s\sum\limits_{l\neq k}\left|\left(\mathbf{WH}\right)_{k,l}\right|^2 + N_0\left(\mathbf{W}\mathbf{W}^H\right)_{k,k}}
\end{equation}

The PPSNR is a good indicator for the error rate performance of MIMO systems, and therefore employed in link adaptation algorithms to predict the uncoded error rate \cite{peng}.  
Since the output of the MMSE detector is Gaussian, the bit error rate of a specific modulation can be calculated by simply plugging the PPSNR value into the AWGN error rate formula of the modulation. The same technique is also used for theoretical derivation of error rate performance in fading channels.

This definition of PPSNR holds if the channel is perfectly known at the receiver. However, in practice, the channel matrix has to be estimated by the receiver, and the estimated channel is inherently noisy in practical systems. We model the estimated channel matrix as
\begin{equation}
\label{estch}
	\widehat{\mathbf{H}}=\mathbf{H}+\Delta\mathbf{H}
\end{equation}
where $\Delta\mathbf{H}$ denotes the estimation error matrix which is uncorrelated with $\mathbf{H}$, and its entries are ZMCSCG with variance $\sigma_e^2$. The quality of channel estimation is captured by $\sigma_e^2$, which can be appropriately estimated depending on the channel estimation method. We assume that each block (packet), that undergoes a specific channel realization, $\mathbf{H}$, observes a different realization of $\Delta\mathbf{H}$ at the receiver. This situation occurs in packet based communication systems like 802.11n where the channel is estimated on a per packet basis.

\subsection{PPSNR derivation for practical systems}
In this section, we derive the PPSNR for practical MIMO systems which observe channel estimation error.  The receiver uses the estimated channel $\widehat{\mathbf{H}}$ to calculate the MMSE detector as
\begin{equation}	
\label{estmmse}		 
	\widehat{\mathbf{W}}=\left[\left(\mathbf{H}+\Delta\mathbf{H}\right)^{H}\left(\mathbf{H}+\Delta\mathbf{H}\right)+\frac{N_0}{E_s}\mathbf{I}\right]^{-1}\left(\mathbf{H}+\Delta\mathbf{H}\right)^H
\end{equation}
We write the imperfect MMSE solution as $\widehat{\mathbf{W}}=\mathbf{W}+\Delta\mathbf{W}$. Now, the MMSE estimate of the signal vector becomes
\begin{equation}
\label{estsignal}		
	\tilde{\mathbf{x}}=\left(\mathbf{W}+\Delta\mathbf{W}\right)\mathbf{y}=\underbrace{\mathbf{W}\mathbf{H}\mathbf{x}}_{\mbox{\footnotesize{signal}}}+\underbrace{\Delta\mathbf{W}\mathbf{H}\mathbf{x}+\mathbf{W}\mathbf{n}+\Delta\mathbf{W}\mathbf{n}}_{\mbox{\footnotesize{post-detection noise}}}
\end{equation}
We observe that there are additional interference and noise terms caused by $\Delta\mathbf{W}$, and denote the post detection noise as $\hat{\mathbf{n}}=\Delta\mathbf{W}\mathbf{H}\mathbf{x}+\mathbf{W}\mathbf{n}+\Delta\mathbf{W}\mathbf{n}$. With this definition for the post detection noise, the PPSNR of the $k^{th}$ spatial stream in the presence of channel estimation error can be expressed as
\begin{equation}
\label{estppsnr}
	\widetilde{\gamma_k}=\frac{E_s\left|\left(\mathbf{WH}\right)_{k,k}\right|^2}{E_s\sum\limits_{l\neq k}\left|\left(\mathbf{WH}\right)_{k,l}\right|^2 + \left(E\left[\hat{\mathbf{n}}\hat{\mathbf{n}}^H\right]\right)_{k,k}}
\end{equation}
where we replaced the original noise covariance in \eqref{ppsnr} with the covariance of $\hat{\mathbf{n}}$, which is calculated as 
\begin{IEEEeqnarray}{rCl}
\label{cov}
E\left[\hat{\mathbf{n}}\hat{\mathbf{n}}^H\right]&=&E\left[\Delta\mathbf{W}\mathbf{H}\mathbf{x}\mathbf{x}^H\mathbf{H}^H\Delta\mathbf{W}^H\right]+E\left[\mathbf{W}\mathbf{n}\mathbf{n}^H\mathbf{W}^H\right]\IEEEnonumber\\
	&&+\:E\left[\mathbf{W}\mathbf{n}\mathbf{n}^H\Delta\mathbf{W}^H\right]+E\left[\Delta\mathbf{W}\mathbf{n}\mathbf{n}^H\mathbf{W}^H\right]\IEEEnonumber\\
	&&+\:E\left[\Delta\mathbf{W}\mathbf{n}\mathbf{n}^H\Delta\mathbf{W}^H\right]
\end{IEEEeqnarray}
In order to calculate the terms in \eqref{cov}, we need to first derive $\Delta\mathbf{W}$. For small $\sigma_e^2$, the $\Delta\mathbf{H}^H\Delta\mathbf{H}$ term in \eqref{estmmse} becomes negligible compared to others. Hence, we can rewrite \eqref{estmmse} as
\begin{equation}	
\label{estmmse1}		  
	\widehat{\mathbf{W}}\cong\left[\mathbf{H}^H\mathbf{H}+\frac{N_0}{E_s}\mathbf{I}+\mathbf{H}^H\Delta\mathbf{H}+\Delta\mathbf{H}^H\mathbf{H}\right]^{-1}\left(\mathbf{H}+\Delta\mathbf{H}\right)^H
\end{equation}
which can be further simplified using the matrix approximation $\left(\mathbf{P}+\epsilon^2\mathbf{Q}\right)^{-1}\cong\mathbf{P}^{-1}-\epsilon^2\mathbf{P}^{-1}\mathbf{Q}\mathbf{P}^{-1}$ for small $\epsilon^2$. Let us also define $\mathbf{K}=\left(\mathbf{H}^H\mathbf{H}+\frac{N_0}{E_s}\mathbf{I}\right)^{-1}$ for brevity and simplify \eqref{estmmse1} as
\begin{IEEEeqnarray}{rCl}		
\label{estmmse2}		  
	\widehat{\mathbf{W}}&\cong&\left[\mathbf{K}-\mathbf{K}\left(\mathbf{H}^H\Delta\mathbf{H}+\Delta\mathbf{H}^H\mathbf{H}\right)\mathbf{K}\right]\left(\mathbf{H}+\Delta\mathbf{H}\right)^H\\
	&=&\mathbf{K}\mathbf{H}^H-\mathbf{K}\left(\mathbf{H}^H\Delta\mathbf{H}+\Delta\mathbf{H}^H\mathbf{H}\right)\mathbf{K}\mathbf{H}^H\IEEEnonumber\\	&&+\:\mathbf{K}\Delta\mathbf{H}^H-\underbrace{\mathbf{K}\left(\mathbf{H}^H\Delta\mathbf{H}+\Delta\mathbf{H}^H\mathbf{H}\right)\mathbf{K}\Delta\mathbf{H}^H}_{\mbox{\footnotesize{small compared to the other terms}}}
\end{IEEEeqnarray}
Finally the desired error matrix becomes
\begin{equation}	
\label{deltaW}		  
	\Delta\mathbf{W}\cong-\mathbf{K}\left(\mathbf{H}^H\Delta\mathbf{H}+\Delta\mathbf{H}^H\mathbf{H}\right)\mathbf{K}\mathbf{H}^H+\mathbf{K}\Delta\mathbf{H}^H.
\end{equation}
Using the above approximation, we can now calculate the terms in \eqref{cov}. We first note that the third and fourth terms in \eqref{cov} are zero since $E\left[\Delta\mathbf{W}\right]\cong0$. The second terms is $E\left[\mathbf{W}\mathbf{n}\mathbf{n}^H\mathbf{W}^H\right]=N_0\mathbf{W}\mathbf{W}^H$, and the first term becomes $E\left[\Delta\mathbf{W}\mathbf{H}\mathbf{x}\mathbf{x}^H\mathbf{H}^H\Delta\mathbf{W}^H\right]=E_sE\left[\Delta\mathbf{W}\mathbf{H}\mathbf{H}^H\Delta\mathbf{W}^H\right]$. Below, we calculate the first and last terms in \eqref{cov} by plugging the error matrix \eqref{deltaW} into \eqref{cov}. 
\begin{IEEEeqnarray}{rCl}		
\label{term1incov}	  
\lefteqn{E\left[\Delta\mathbf{W}\mathbf{H}\mathbf{H}^H\Delta\mathbf{W}^H\right] }\IEEEnonumber\\ &\cong&E\left[\mathbf{K}\mathbf{H}^H\Delta\mathbf{H}\mathbf{K}\mathbf{H}^H\mathbf{H}\mathbf{H}^H\mathbf{H}\mathbf{K}^H\Delta\mathbf{H}^H\mathbf{H}\mathbf{K}^H\right]\IEEEnonumber\\
&&+\:E\left[\mathbf{K}\mathbf{H}^H\Delta\mathbf{H}\mathbf{K}\mathbf{H}^H\mathbf{H}\mathbf{H}^H\mathbf{H}\mathbf{K}^H\mathbf{H}^H\Delta\mathbf{H}\mathbf{K}^H\right]\IEEEnonumber\\
&&-\:E\left[\mathbf{K}\mathbf{H}^H\Delta\mathbf{H}\mathbf{K}\mathbf{H}^H\mathbf{H}\mathbf{H}^H\Delta\mathbf{H}\mathbf{K}^H\right]\IEEEnonumber\\
&&+\:E\left[\mathbf{K}\Delta\mathbf{H}^H\mathbf{H}\mathbf{K}\mathbf{H}^H\mathbf{H}\mathbf{H}^H\mathbf{H}\mathbf{K}^H\Delta\mathbf{H}^H\mathbf{H}\mathbf{K}^H\right]\IEEEnonumber\\
&&+\:E\left[\mathbf{K}\Delta\mathbf{H}^H\mathbf{H}\mathbf{K}\mathbf{H}^H\mathbf{H}\mathbf{H}^H\mathbf{H}\mathbf{K}^H\mathbf{H}^H\Delta\mathbf{H}\mathbf{K}^H\right]\IEEEnonumber\\
&&-\:E\left[\mathbf{K}\Delta\mathbf{H}^H\mathbf{H}\mathbf{K}\mathbf{H}^H\mathbf{H}\mathbf{H}^H\Delta\mathbf{H}\mathbf{K}^H\right]\IEEEnonumber\\
&&-\:E\left[\mathbf{K}\Delta\mathbf{H}^H\mathbf{H}\mathbf{H}^H\mathbf{H}\mathbf{K}^H\Delta\mathbf{H}^H\mathbf{H}\mathbf{K}^H\right]\IEEEnonumber\\
&&-\:E\left[\mathbf{K}\Delta\mathbf{H}^H\mathbf{H}\mathbf{H}^H\mathbf{H}\mathbf{K}^H\mathbf{H}^H\Delta\mathbf{H}\mathbf{K}^H\right]\IEEEnonumber\\
&&+\:E\left[\mathbf{K}\Delta\mathbf{H}^H\mathbf{H}\mathbf{H}^H\Delta\mathbf{H}\mathbf{K}^H\right]
\end{IEEEeqnarray}
It can be proven that $E\left[\Delta\mathbf{H}\mathbf{A}\Delta\mathbf{H}\right]=E\left[\Delta\mathbf{H}^H\mathbf{A}\Delta\mathbf{H}^H\right]=0$ for any deterministic matrix $\mathbf{A}$. Hence the second, third, fourth and seventh terms in (13) are zero. For the remaining terms we use the fact that $E\left[\Delta\mathbf{H}\mathbf{A}\Delta\mathbf{H}^H\right]=\sigma_e^{2}\tr\left(\mathbf{A}\right)\mathbf{I}$, and obtain
\begin{IEEEeqnarray}{rCl}			  
\lefteqn{E\left[\Delta\mathbf{W}\mathbf{H}\mathbf{H}^H\Delta\mathbf{W}^H\right] }\IEEEnonumber\\ &\cong&\sigma_e^{2}\tr\left(\mathbf{K}\mathbf{H}^H\mathbf{H}\mathbf{H}^H\mathbf{H}\mathbf{K}^H\right)\mathbf{K}\mathbf{H}^H\mathbf{H}\mathbf{K}^H\IEEEnonumber\\
&&+\:\sigma_e^{2}\tr\left(\mathbf{H}\mathbf{K}\mathbf{H}^H\mathbf{H}\mathbf{H}^H\mathbf{H}\mathbf{K}^H\mathbf{H}^H\right)\mathbf{K}\mathbf{K}^H\IEEEnonumber\\
&&-\:\sigma_e^{2}\tr\left(\mathbf{H}\mathbf{K}\mathbf{H}^H\mathbf{H}\mathbf{H}^H\right)\mathbf{K}\mathbf{K}^H\\
&&-\:\sigma_e^{2}\tr\left(\mathbf{H}\mathbf{H}^H\mathbf{H}\mathbf{K}^H\mathbf{H}^H\right)\mathbf{K}\mathbf{K}^H + \sigma_e^{2}\tr\left(\mathbf{H}\mathbf{H}^H\right)\mathbf{K}\mathbf{K}^H\IEEEnonumber
\end{IEEEeqnarray}
Similarly, the last term in \eqref{cov}, $E\left[\Delta\mathbf{W}\mathbf{n}\mathbf{n}^H\Delta\mathbf{W}^H\right]$, can be computed following the same way.
\begin{equation}
E\left[\Delta\mathbf{W}\mathbf{n}\mathbf{n}^H\Delta\mathbf{W}^H\right]=N_0E\left[\Delta\mathbf{W}\Delta\mathbf{W}^H\right]
\end{equation}
\begin{IEEEeqnarray}{rCl}			  
\lefteqn{E\left[\Delta\mathbf{W}\Delta\mathbf{W}^H\right] }\IEEEnonumber\\ &\cong&\sigma_e^{2}\tr\left(\mathbf{K}\mathbf{H}^H\mathbf{H}\mathbf{K}^H\right)\mathbf{K}\mathbf{H}^H\mathbf{H}\mathbf{K}^H\IEEEnonumber\\
&&+\:\sigma_e^{2}\tr\left(\mathbf{H}\mathbf{K}\mathbf{H}^H\mathbf{H}\mathbf{K}^H\mathbf{H}^H\right)\mathbf{K}\mathbf{K}^H\IEEEnonumber\\
&&-\:\sigma_e^{2}\tr\left(\mathbf{H}\mathbf{K}\mathbf{H}^H\right)\mathbf{K}\mathbf{K}^H\IEEEnonumber\\
&&-\:\sigma_e^{2}\tr\left(\mathbf{H}\mathbf{K}^H\mathbf{H}^H\right)\mathbf{K}\mathbf{K}^H + \sigma_e^{2}N_r\mathbf{K}\mathbf{K}^H
\end{IEEEeqnarray}
Finally, we plug $E\left[\hat{\mathbf{n}}\hat{\mathbf{n}}^H\right]$ into \eqref{estppsnr} and obtain the PPSNR in the presence of channel estimation error as \eqref{eqn_dbl_x}.
\begin{figure*}[!t]
\normalsize
\setcounter{MYtempeqncnt}{\value{equation}}
\setcounter{equation}{16}

\begin{equation}
\label{eqn_dbl_x}
\widetilde{\gamma_k}\cong\!\frac{E_s\left|\left(\mathbf{WH}\right)_{k,k}\right|^2}{E_s\sum\limits_{l\neq k}\left|\left(\mathbf{WH}\right)_{k,l}\right|^2+
\left(
\begin{split}
&E_s\sigma_e^{2}\tr\!\left(\mathbf{K}\mathbf{H}^H\mathbf{H}\mathbf{H}^H\mathbf{H}\mathbf{K}^H\right)\mathbf{K}\mathbf{H}^H\mathbf{H}\mathbf{K}^H+E_s\sigma_e^{2}\tr\!\left(\mathbf{H}\mathbf{K}\mathbf{H}^H\mathbf{H}\mathbf{H}^H\mathbf{H}\mathbf{K}^H\mathbf{H}^H\right)\mathbf{K}\mathbf{K}^H  \\[-3pt] &-E_s\sigma_e^{2}\tr\!\left(\mathbf{H}\mathbf{K}\mathbf{H}^H\mathbf{H}\mathbf{H}^H\right)\!\mathbf{K}\mathbf{K}^H\!\!-\!E_s\sigma_e^{2}\tr\!\left(\mathbf{H}\mathbf{H}^H\mathbf{H}\mathbf{K}^H\mathbf{H}^H\right)\!\mathbf{K}\mathbf{K}^H\!\!+\!E_s\sigma_e^{2}\tr\!\left(\mathbf{H}\mathbf{H}^H\right)\!\mathbf{K}\mathbf{K}^H  \\[-3pt]
&+N_0\mathbf{W}\mathbf{W}^H+N_0\sigma_e^{2}\tr\!\left(\mathbf{K}\mathbf{H}^H\mathbf{H}\mathbf{K}^H\right)\mathbf{K}\mathbf{H}^H\mathbf{H}\mathbf{K}^H+N_0\sigma_e^{2}\tr\!\left(\mathbf{H}\mathbf{K}\mathbf{H}^H\mathbf{H}\mathbf{K}^H\mathbf{H}^H\right)\mathbf{K}\mathbf{K}^H \\[-3pt]
&-N_0\sigma_e^{2}\tr\!\left(\mathbf{H}\mathbf{K}\mathbf{H}^H\right)\mathbf{K}\mathbf{K}^H-N_0\sigma_e^{2}\tr\!\left(\mathbf{H}\mathbf{K}^H\mathbf{H}^H\right)\mathbf{K}\mathbf{K}^H+N_0\sigma_e^{2}N_r\mathbf{K}\mathbf{K}^H 
\end{split}
\right)_{\!k,k}}
\end{equation}
\setcounter{equation}{\value{MYtempeqncnt}}
\hrulefill
\end{figure*}

The BER of the system in the presence of channel estimation error can be found simply by plugging $\widetilde{\gamma_k}$ as the symbol SNR into the AWGN BER formulas. For example, the BER of $k^{th}$ stream for BPSK is $P_b^k=Q\left(\sqrt{2\widetilde{\gamma_k}}\right)$, and $P_b^k=\frac{3}{4}Q\left(\sqrt{\frac{\widetilde{\gamma_k}}{5}}\right)+\frac{1}{2}Q\left(3\sqrt{\frac{\widetilde{\gamma_k}}{5}}\right)-\frac{1}{4}Q\left(5\sqrt{\frac{\widetilde{\gamma_k}}{5}}\right)$ for gray-coded 16QAM.
\section{Results}
In order to test the performance of the analysis, we simulated transmission of thousands of packets through uncorrelated Rayleigh flat fading channels. For each SNR point on the BER plots, we randomly generate 1000 i.i.d. realizations of the channel matrix $\mathbf{H}$. For each specific realization of the channel, we transmit 500 packets each of which carries 2000 information symbols. We perform channel estimation for each packet as explained below in \textit{Case 1}.

\textit{Case 1:} In our simulations, we employed the maximum likelihood (ML) channel estimation (CE) algorithm, in which the channel estimate is obtained via training symbols that are known to the receiver. During the training phase, the $N_t \times N_{tr}$ training matrix $\mathbf{X}_{tr}$ is transmitted where $N_{tr}\geq N_t$ is the number of training symbols. The $N_r \times N_{tr}$ received signal is $\mathbf{Y}_{tr}=\mathbf{H}\mathbf{X}_{tr}+\mathbf{W}$ where $\mathbf{W}$ is the $N_r \times N_{tr}$ noise matrix. Then, the ML estimate of the channel is given as \cite{channel_est}
\setcounter{equation}{17}
\begin{equation}
\widehat{\mathbf{H}}=\mathbf{Y}_{tr}\mathbf{X}_{tr}^H\left(\mathbf{X}_{tr}\mathbf{X}_{tr}^H\right)^{-1}
\end{equation}
It was shown that the optimal training signal has the property of $\mathbf{X}_{tr}\mathbf{X}_{tr}^H=E_sN_{tr}\mathbf{I}$. When this orthogonal training signal is employed, the entries of $\Delta\mathbf{H}$ are i.i.d. with $CN(0,\sigma_e^2)$, and the channel estimation noise variance\footnote{$\sigma_e^2$ can also be defined as $\sigma_e^2=\frac{N_t}{N_{tr}E_s/N_0}$ depending on SNR definition.} is $\sigma_e^2=\frac{1}{N_{tr}E_s/N_0}$ \cite{channel_est}. The estimation error in this case is caused by the AWGN in this case. 

The following training signal, which is taken from 802.11n standard \cite{n80211}, was employed in the simulations. $\mathbf{X}_{tr}=\sqrt{E_s}\ddot{\mathbf{P}}$ where $\ddot{\mathbf{P}}$ is the submatrix formed by first $N_t$ rows and first $N_{tr}$ columns of the bigger matrix\footnote{The $\mathbf{P}$ matrix here is for maximum of 4 spatial streams since the standard supports up to 4 streams. $N_{tr}=4$ for $N_t\geq3$.} $\mathbf{P}$, i.e. $\ddot{\mathbf{P}}=\mathbf{P}\left[1:N_t;1:N_{tr}\right]$.
\begin{equation}
\mathbf{P}=\left[{\begin{array}{cccc}
1&-1&1&1\\
1&1&-1&1\\
1&1&1&-1\\
-1&1&1&1\\
\end{array}}\right]
\end{equation}
It should be noted that with this choice of the training matrix, the ML channel estimation at the receiver becomes a very simple operation since the matrix inversion, $\left(\mathbf{X}_{tr}\mathbf{X}_{tr}^H\right)^{-1}$, is now a trivial operation.

\begin{figure}[!t]
\centering
		\includegraphics[trim = 10mm 0mm 8mm 6mm, width=3.4in]{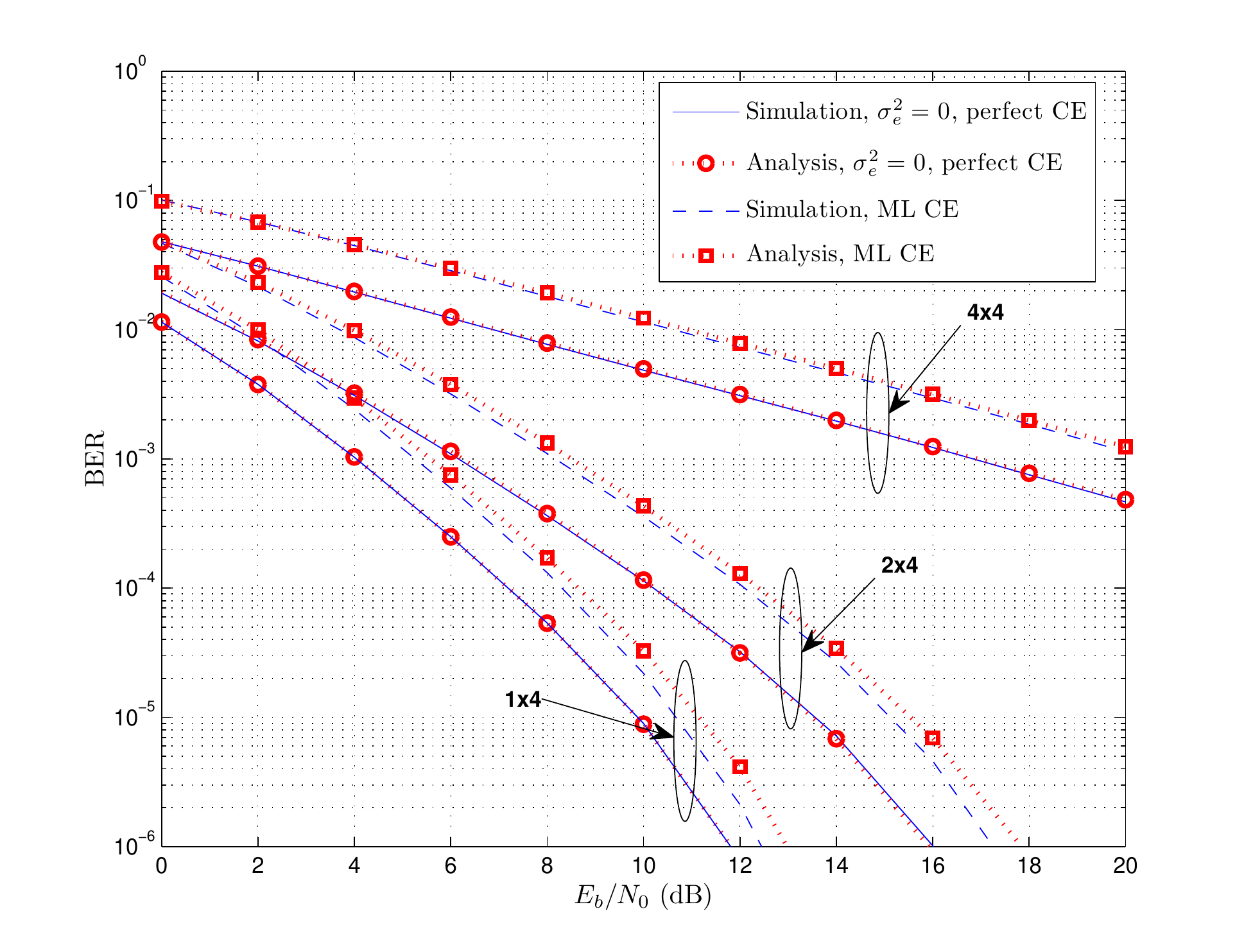}
\caption{BER for BPSK. $E_b/N_0$ is per tx antenna.}
\label{bpsk_all_latex3}
\end{figure}

\begin{figure}[!t]
\centering
		\includegraphics[trim = 10mm 0mm 8mm 6mm,width=3.4in]{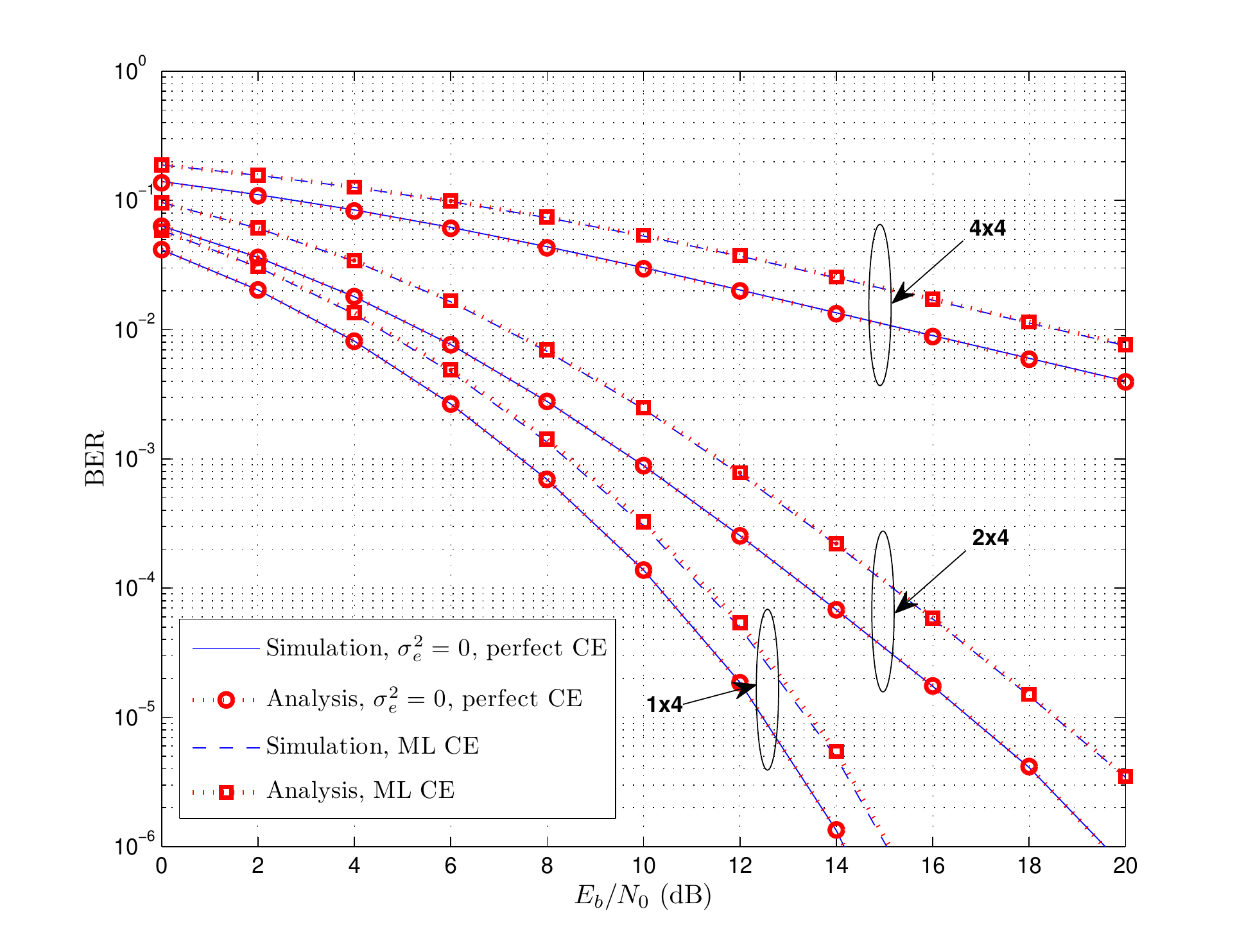}
\caption{BER for 16QAM. $E_b/N_0$ is per tx antenna.}
\label{qam16_all_latex3}
\end{figure}

We present the simulation results for BPSK in Fig. \ref{bpsk_all_latex3} and 16QAM in Fig. \ref{qam16_all_latex3} with $1\times4$, $2\times4$, $4\times4$ MIMO configurations. $N_{tr}=4$ is used in all the simulations. The case of $\sigma_e^2=0$, i.e. perfect channel estimation, is also included in the results.

For each channel instance, analytical BER results are obtained by using the PPSNR derived in the previous section. Then, these BERs are averaged over all realizations of the channel.

First thing to notice in Fig. \ref{bpsk_all_latex3} and Fig. \ref{qam16_all_latex3} is that for $\sigma_e^2=0$, simulation and analysis curves exactly match. Performance is significantly degraded for the systems experiencing channel estimation errors. This is particularly evident for the $4\times4$ configurations. 

As it can be seen in Fig. \ref{bpsk_all_latex3} and Fig. \ref{qam16_all_latex3}, our analysis gives a very tight approximation of the real performance. For BPSK $4\times4$, and all of the 16QAM configurations the analysis results exactly match the simulated performances. For BPSK $1\times4$, and $2\times4$ configurations the analysis results are upper-bounds to the real performance at high SNR, however, they are still very close to the real performances. The analysis results become tighter for higher order modulations and higher order MIMO configurations. This is because of the fact that the Gaussian assumption, which is made for the post-detection noise, is more valid at higher order modulations and MIMO configurations. At low SNRs, the total post detection noise $\hat{\mathbf{n}}$ is dominated by the additive white Gaussian noise component $\mathbf{n}$ therefore the assumption is valid even for lower configurations. However, at high SNRs the residual interference components from other spatial streams becomes dominant and $\hat{\mathbf{n}}$ is loosely approximated as Gaussian for lower order constellations and MIMO configurations.

It is interesting to note that in contrast to the results obtained for ZF detector by \cite{wang}, we do not observe any error floor on the performance. This is due to the fact that the channel estimation error variance $\sigma_e^2$ for ML estimation gets smaller as SNR increases. This is the situation that occurs in practical packet based or bursty communication systems where the channel estimation is performed for every packet prior to data detection, and hence experiences the same noise variance as the data transmission. Therefore the channel estimation quality is dependent on the SNR. On the other hand, error floors are observed in \cite{wang} because of the assumption that $\sigma_e^2$ remains constant independent of the SNR. This case is investigated below in \textit{Case 2}.

\begin{figure}[!t]
\centering
		\includegraphics[trim = 10mm 0mm 8mm 6mm,width=3.4in]{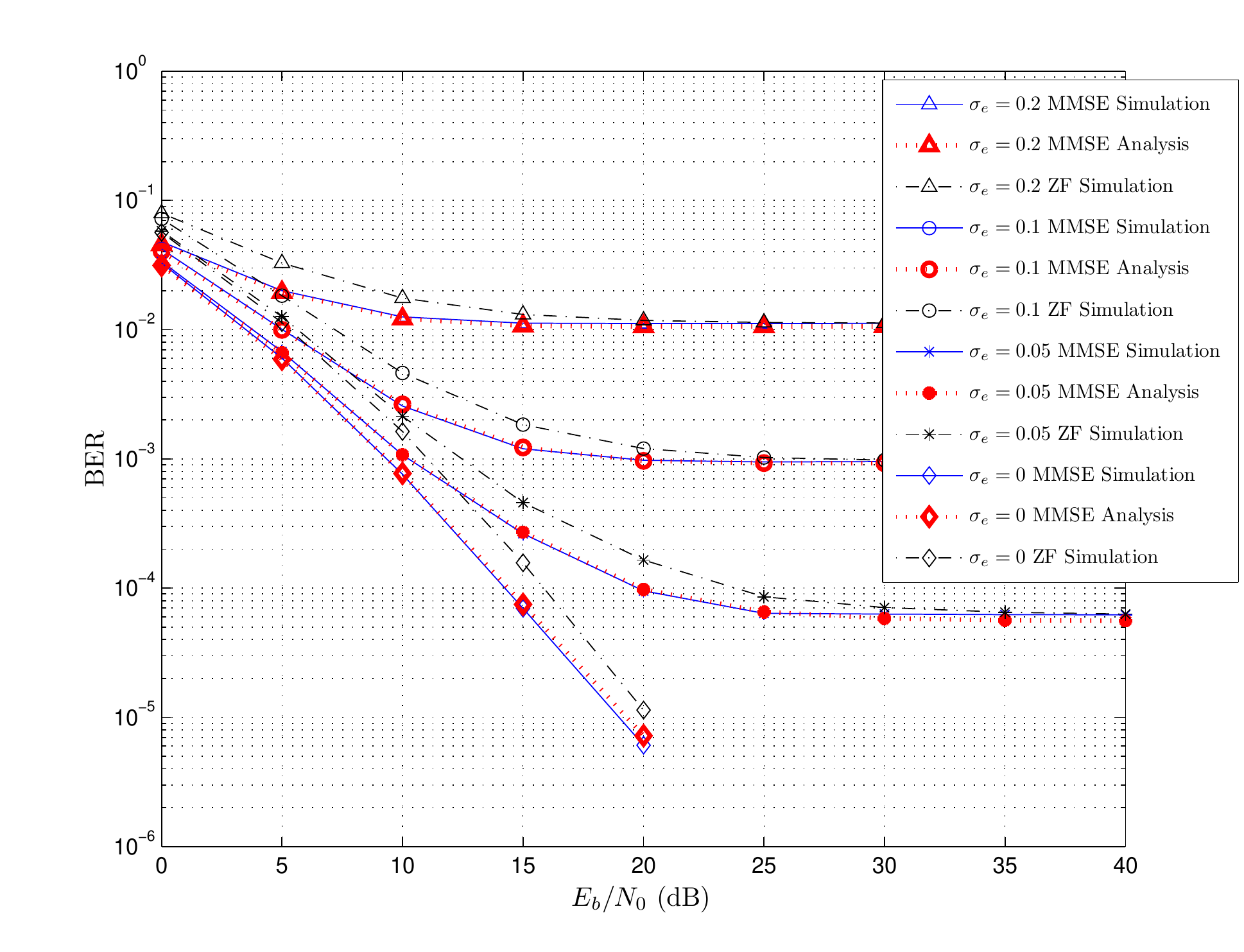}
\caption{BER for $4\times 5$ QPSK for different $\sigma_e$ values. $\sigma_e=0$ corresponds to perfect CE. ZF curves are included for comparison purposes with \cite{wang}. $E_b/N_0$ is per tx antenna. These parameter values are taken from Fig. 2 of \cite{wang} for comparison purposes.}
\label{qpsk}
\end{figure}

\textit{Case 2:} In addition to ML channel estimation results, we also performed simulations with constant $\sigma_e$. Unlike the first case, the channel estimation quality is independent of the SNR. This situation might arise either when there is a ready channel estimate to be used by the receiver formed elsewhere with a different additive noise variance, or the channel estimation is outdated and the major error in the channel estimation comes from the mobility changes in the channel.

In Fig. \ref{qpsk}, the BER performance of a QPSK $4\times 5$ system is investigated for $\sigma_e=5\%,10\%,20\%$ using the estimation error model in (4). Each packet observes a different realization of the random matrix $\Delta\mathbf{H}$ with the designated variance $\sigma_e^2$. As expected, we observe error floor in the performance due to the constant estimation error variance as in the ZF detector case studied in \cite{wang}. More importantly, these error floors are the same as the ones observed by ZF detector because of the fact that the MMSE and ZF detectors exhibit the same behaviour at asymptotically high SNR. The simulation results in this case also agree with the analysis.

\section{Conclusion}
In this letter, we presented the analysis of post-processing SNR for practical MIMO MMSE receivers which experience imperfect channel estimation. Performance of MMSE receivers in the presence of channel estimation error is investigated and shown to be accurately estimated via analytical results. We verified the tightness of the analytical results via simulations. 

Besides the theoretical contributions, we believe that our closed form PPSNR expression can be useful for link adaptation purposes in real MIMO systems. There exist link adaptation algorithms \cite{peng,wcnc12} based on PPSNR, however perfect CSI is always assumed which might lead to incorrect prediction of the throughput. More accurate prediction can be achieved using the results presented in this paper.


%




\ifCLASSOPTIONcaptionsoff
  \newpage
\fi

\end{document}